\documentclass[cameraready]{Interspeech}

\title{Stabilizing Short Duration Speaker Verification through Neural Re-scoring with Hybrid Enrollment}

\author[affiliation={1}, orcid=0009-0005-1034-9972]{Zhiqi}{Ai}
\author[affiliation={1}, orcid=0009-0007-8540-3990]{Han}{Cheng}
\author[affiliation={1}, orcid=0000-0002-5393-0258]{Shiyi}{Mu}
\author[affiliation={3}, orcid=0000-0002-9629-6111]{Zhiyong}{Chen}
\author[affiliation={1}, orcid=0000-0002-7966-2992, correspondingauthor]{Yongjin}{Zhou}
\author[affiliation={2}, orcid=0000-0003-1905-6269, correspondingauthor]{Shugong}{Xu}

\address{
    $^1$ Shanghai University, China \\
    $^2$ Xi’an Jiaotong-Liverpool University, China \\
    $^3$ Hithink RoyalFlush AI Research Institute, China
}

\email{\{aizhiqi-work, yjzhou\}@shu.edu.cn, shugong.xu@xjtlu.edu.cn}


\usepackage{comment}
\usepackage{multirow}
\usepackage[table,xcdraw]{xcolor}

\usepackage{booktabs}   
\usepackage{multirow}   
\usepackage{makecell}   
\usepackage[table]{xcolor} 
\usepackage{graphicx}
\usepackage{amsmath}

\begin{document}

\maketitle

\begin{abstract}

Short-duration speaker verification (SDSV) is crucial for personalized keyword spotting, where test utterances are typically shorter than three seconds. Limited speech duration results in unstable speaker representations and increased sensitivity to noise and phoneme variations, thereby degrading performance. To investigate this issue, we construct VoxPhrase, a large-scale SDSV corpus automatically segmented from the VoxCeleb dataset. Our analysis shows that text-dependent (TD) enrollment is constrained by duration and yields unstable speaker representations. In contrast, although text-independent (TI) enrollment introduces content mismatch, its representations become more stable as the enrollment duration increases. Accordingly, we propose a hybrid-enrollment neural re-scoring framework that combines TD and TI enrollment and performs frame-level comparison via parallel cross-attention. Experiments on VoxPhrase demonstrate consistent improvements across multiple speaker models.
        
\end{abstract}

\section{Introduction}

\begin{figure}[t]
  \centering
  \includegraphics[width=0.9\linewidth]{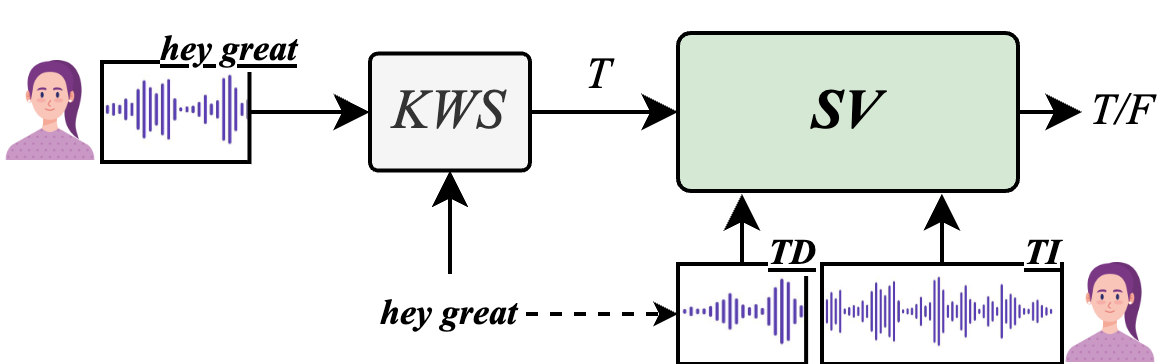}
  \caption{Schematic of the short-duration speaker verification task for custom phrases. The input speech is first filtered by a custom keyword spotting module, and the detected target utterance is then verified against text-dependent (TD) or text-independent (TI) enrollment speech.}
  \label{fig:f1}
\end{figure}

With the widespread adoption of smart devices and conversational terminals, user-defined keyword spotting (UDKWS) \cite{baseline_cmcd, cacd, mfa-kws, ai24_interspeech, ds-kws, dma-kws} has been widely deployed. In practical systems, speaker verification (SV) \cite{ecapa-tdnn, cam++, eres2net} is typically performed after detecting the target phrase to ensure security. As illustrated in Figure~\ref{fig:f1}, verification is conducted on custom phrase-bounded speech segments, which are usually short (often less than three seconds). This gives rise to short-duration speaker verification (SDSV) \cite{sdsv2020}. In this setting, the system must determine speaker identity from extremely brief speech segments with limited speaker-discriminative information, substantially increasing verification difficulty.

Modern SV systems typically rely on fixed-dimensional speaker embeddings with similarity-based backends. Models such as ECAPA-TDNN \cite{ecapa-tdnn} and CAM++ \cite{cam++} have become mainstream foundation models and are widely integrated into open-source toolkits like 3D-Speaker \cite{3dspeaker}. These methods perform reliably in text-independent, long-utterance scenarios. However, in SDSV settings, speaker representations become more susceptible to lexical content. Text-dependent (TD) enrollment ensures content consistency but is constrained by short duration and reduced speaker information, whereas text-independent (TI) enrollment provides richer speaker evidence yet introduces content mismatch with phrase-bounded test speech.

Existing studies on SDSV remain limited and are primarily conducted on small-scale TD corpora, such as DeepMine \cite{deepmine} and Qcomm \cite{9004014}. Some works \cite{sdsv2020, lozanodiez20_interspeech} fine-tune models on predefined target-phrase datasets to enhance embedding discrimination under phrase constraints. Other studies \cite{chen2024optimizing, chen25_interspeech, 10832359} build upon strong TI foundation models and introduce lightweight adaptation mechanisms (e.g., adapters) under registered-speaker and known-phrase settings. These approaches typically assume fixed target phrases and predefined user sets. In contrast, in UDKWS scenarios, target phrases are freely specified by users, making predefined-vocabulary modeling impractical, and research in this setting remains limited.

To better understand SDSV under user-defined settings, we construct VoxPhrase, a large-scale SDSV corpus automatically segmented from the VoxCeleb \cite{vox1, vox2} dataset. VoxPhrase enables systematic investigation under customizable enrollment conditions. Our analysis shows that TD enrollment, while content-consistent, is constrained by limited duration and produces unstable speaker representations. In contrast, TI enrollment introduces lexical mismatch but yields increasingly stable representations as enrollment duration grows. Based on this analysis, we propose a hybrid-enrollment neural re-scoring framework that jointly leverages TD and TI enrollment speech. The proposed verifier performs frame-level matching via parallel cross-attention to refine similarity estimation. Extensive experiments on VoxPhrase demonstrate consistent improvements across multiple speaker embedding backbones, particularly under short-duration conditions.

\begin{figure*}[!htp]
  \centering
  \includegraphics[width=0.9\linewidth]{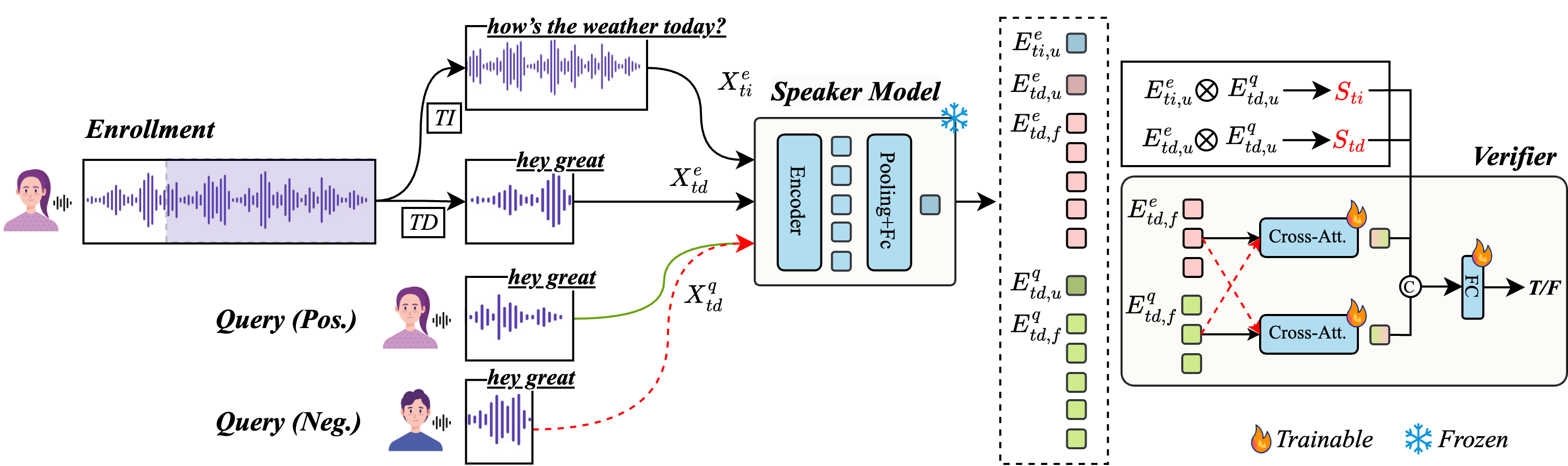}
  \caption{Overview of the proposed framework for short-duration speaker verification. The enrollment stage adopts a hybrid registration scheme that combines text-dependent (TD) and text-independent (TI) utterances. Enrollment and query speech are first processed by a frozen speaker model to extract both frame-level and utterance-level embeddings. The verifier then applies a parallel cross-attention module to model interactions between enrollment and query frame-level speaker features, and fuses them with the utterance-level similarity to produce the final verification score (T/F).}
  \label{fig:overview}
\end{figure*}

\section{Proposed Framework}
\label{sec:method}

As illustrated in Figure~\ref{fig:overview}, the proposed framework consists of two key components: (1) a frozen speaker backbone for feature extraction, and (2) a trainable neural verifier for similarity modeling. To enhance robustness under short-duration conditions, the system adopts a hybrid enrollment strategy that combines text-independent (TI) and text-dependent (TD) speech. For each target speaker, TI enrollment speech $X^e_{ti}$ captures stable speaker identity traits, while TD enrollment speech $X^e_{td}$ preserves phrase-consistent characteristics. During inference, the query speech $X^q_{td}$ is always text-dependent. By integrating the global identity information provided by hybrid enrollment with fine-grained frame-level matching between enrollment and query speech, the system achieves robust short-duration speaker verification.

\subsection{Speaker Feature Extraction}

All enrollment and query utterances are processed by a frozen pretrained speaker model, which extracts utterance-level and frame-level representations. Specifically, TI enrollment speech $X^e_{ti}$ produces the utterance-level representation $\mathbf{E}^{e}_{ti,u}$ and the frame-level representation $\mathbf{E}^{e}_{ti,f}$. TD enrollment speech $X^e_{td}$ produces $\mathbf{E}^{e}_{td,u}$ and $\mathbf{E}^{e}_{td,f}$, while TD query speech $X^q_{td}$ produces $\mathbf{E}^{q}_{td,u}$ and $\mathbf{E}^{q}_{td,f}$. The backbone remains frozen to preserve the pretrained speaker discrimination capability, while the downstream verifier is optimized for task-specific similarity modeling.

\subsection{Neural Verifier}
\label{sec:c0}

The neural verifier models speaker similarity from both utterance-level (global) and frame-level (local) representations.
\subsubsection{Global Similarity}
Two cosine similarity scores are computed from utterance-level representations:
\begin{equation}
    S_{ti} = \cos(\mathbf{E}^{e}_{ti,u}, \mathbf{E}^{q}_{td,u}),
    \quad
    S_{td} = \cos(\mathbf{E}^{e}_{td,u}, \mathbf{E}^{q}_{td,u})
\end{equation}
where $S_{ti}$ captures speaker identity consistency from TI enrollment speech, and $S_{td}$ reflects phrase-consistent similarity under the TD condition.

\subsubsection{Parallel Cross-Attention}

To capture fine-grained temporal correspondence, a shared parallel cross-attention module operates on frame-level representations. Enrollment-to-query attention is defined as:
\begin{equation}
    \tilde{\mathbf{Z}}_e =
    \text{Cross-Att.}(
    \mathbf{Q}=\mathbf{E}^{e}_{td,f},
    \mathbf{K}=\mathbf{E}^{q}_{td,f},
    \mathbf{V}=\mathbf{E}^{q}_{td,f}),
\end{equation}
and query-to-enrollment attention is:
\begin{equation}
    \tilde{\mathbf{Z}}_q =
    \text{Cross-Att.}(
    \mathbf{Q}=\mathbf{E}^{q}_{td,f},
    \mathbf{K}=\mathbf{E}^{e}_{td,f},
    \mathbf{V}=\mathbf{E}^{e}_{td,f}).
\end{equation}
The bidirectional cross-attention design enables more robust modeling of temporal interactions between enrollment and query speech, particularly in short-duration scenarios where alignment is inherently unstable \cite{kim2024bridging}.

The outputs are temporally max-pooled and concatenated:
\begin{equation}
    \mathbf{h}_f =
    \left[
    \max(\tilde{\mathbf{Z}}_e)
    \parallel
    \max(\tilde{\mathbf{Z}}_q)
    \right].
\end{equation}

\subsection{Fusion and Decision}
Global and local evidence are fused through a lightweight MLP:
\begin{equation}
    S = \sigma\big(\mathcal{F}(\mathbf{h}_f, S_{ti}, S_{td})\big),
\end{equation}
where $\sigma(\cdot)$ denotes the sigmoid function. 

The model is trained using binary cross-entropy:
\begin{equation}
    \mathcal{L}
    =
    - \big[
    y \log S + (1-y)\log(1-S)
    \big],
\end{equation}
where $y \in \{0,1\}$ indicates whether the enrollment and query utterances belong to the same speaker.

\section{VoxPhrase Construction and Evaluation}

\subsection{Construction Pipeline}
\label{sec:c1}

\begin{figure}[t]
  \centering
  \includegraphics[width=0.9\linewidth]{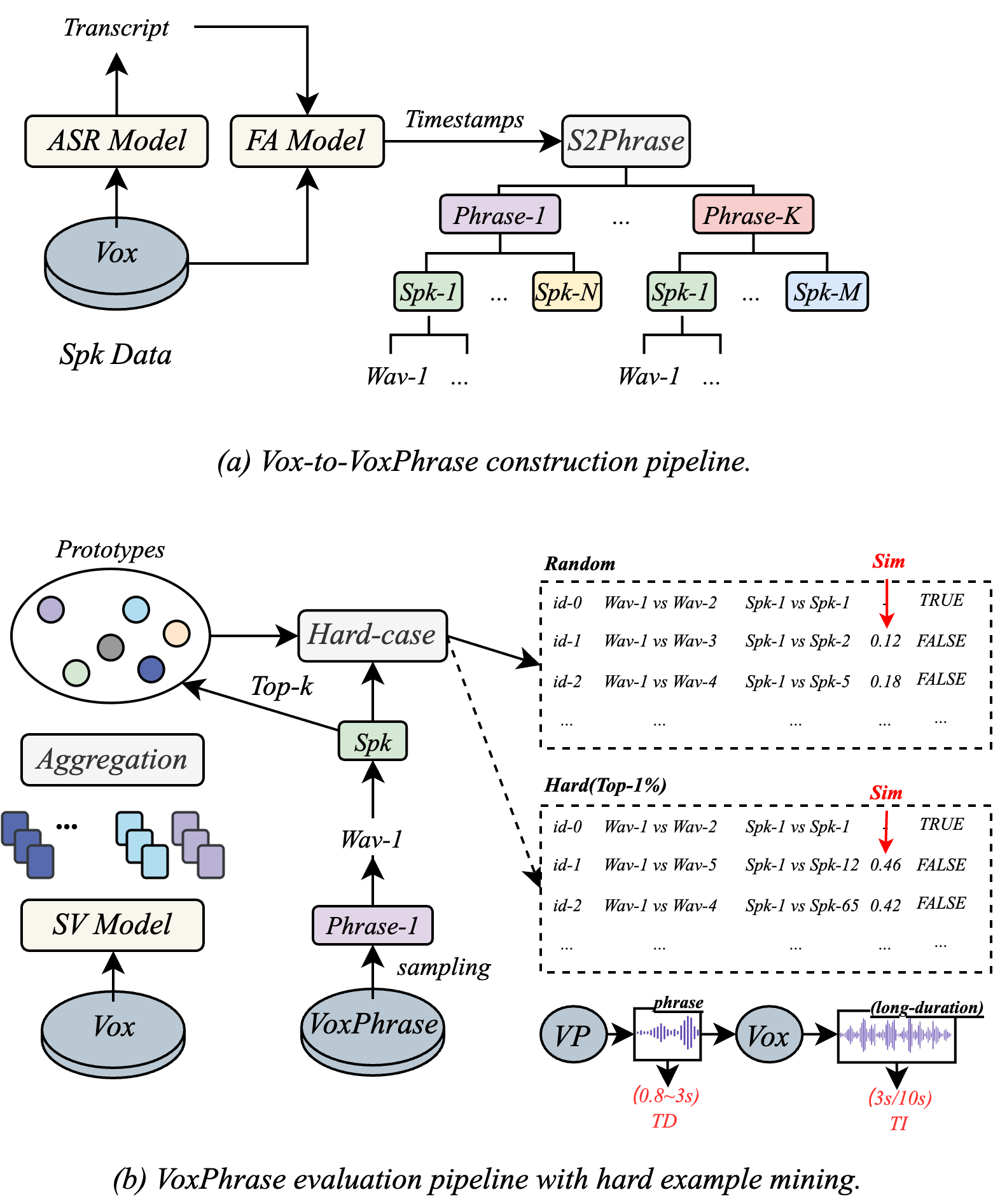}
  \caption{Overview of the VoxPhrase dataset construction and evaluation pipeline.}
  \label{fig:data}
\end{figure}

As illustrated in Figure~\ref{fig:data}(a), to construct the VoxPhrase dataset for short-duration text-dependent speaker verification, we start from the original VoxCeleb corpus. Following commonly adopted phrase segmentation strategies in custom keyword spotting \cite{baseline_cmcd, ai24_interspeech, ds-kws, dma-kws}, we first apply an automatic speech recognition (ASR) \cite{rekesh2023fast} model to obtain transcripts for each utterance. A forced alignment (FA) \cite{baevski2020wav2vec} model is then employed to generate accurate word- or phrase-level timestamps. Based on the alignment results, an S2Phrase script is used to segment long-form speech into phrase-level units. During segmentation, low-quality alignment results are filtered out, and phrase durations are constrained within 0.8--3 seconds to ensure consistency and reliability. Each extracted phrase segment is associated with its corresponding speaker identity and waveform. Through this process, long utterances in VoxCeleb are transformed into structured phrase-level speaker data, enabling short-duration text-dependent speaker verification. 

\subsection{Evaluation with Hard Example Mining}
\label{sec:c2}

As illustrated in Figure~\ref{fig:data}(b), to systematically evaluate performance under short-duration conditions, we design an evaluation pipeline incorporating hard example mining. Phrase-level samples are first grouped by speaker identity, and a pretrained speaker verification (SV) model is used to extract embeddings and construct speaker prototypes via aggregation. Inter-speaker similarity scores are then computed, and highly similar but identity-mismatched speakers are selected as hard negatives. These hard negative pairs, together with genuine pairs, form the final evaluation set. In addition, to assess performance under text-independent (TI) enrollment, speech segments are randomly sampled from the original Vox data based on phrase-level timestamps and used as TI enrollment data, enabling evaluation under text-independent registration settings.

\section{Experiments}

\subsection{Experimental Setup}

The statistics of VoxPhrase are summarized in Table~\ref{tab:vp_stats}. 
The training set is constructed from Vox2-dev \cite{vox2} with 5,994 speakers. 
Evaluation includes four subsets: 
\textbf{Eval-1} (Vox1, 1,251 speakers) \cite{vox1} and 
\textbf{Eval-2} (Vox2-test, 118 speakers) for large-scale evaluation; 
\textbf{Eval-3} and \textbf{Eval-4} are built from Deepmine \cite{deepmine} short phrases, namely 
``ok google'' ($\sim$2\,s) and 
``my voice is my password'' ($\sim$3\,s), 
with remaining utterances used as TI enrollment candidates for 
out-of-distribution evaluation. 
Following Section~\ref{sec:c2}, we further analyze the hard-case subset. 
Trials are constructed under different anchor selection strategies 
(top-1\%, top-5\%, top-10\%, and random) to study lexical consistency effects. 
Performance is measured using Equal Error Rate (EER, \%), 
and average EER across anchor settings is reported.

\subsection{Training Setup}

We adopt three open-source speaker models: ECAPA-TDNN \cite{ecapa-tdnn}\footnote{\url{https://modelscope.cn/models/iic/speech_ecapa-tdnn_sv_en_voxceleb_16k}}, CAM++ \cite{cam++}\footnote{\url{https://modelscope.cn/models/iic/speech_campplus_sv_en_voxceleb_16k}}, and ERes2Net-L \cite{eres2net}\footnote{\url{https://modelscope.cn/models/iic/speech_eres2net_large_sv_en_voxceleb_16k}}. 
All models are pretrained on Vox2 and kept frozen during training. 
Frame-level and utterance-level embeddings are extracted from the speaker models. 
On top of the frozen encoder, we train a lightweight verifier (Section~\ref{sec:c0}) 
consisting of a linear projection layer and a symmetric cross-attention module 
(8 heads, hidden dimension 128) to produce the matching score. 
The verifier is optimized using binary cross-entropy over target and non-target trials. 
Training is conducted on a single RTX~4090 GPU with batch size 256 for 25k steps.

\begin{table}[]
\centering
\renewcommand{\arraystretch}{1.35}
\caption{Training and evaluation protocol for short-duration speaker verification on VoxPhrase.}
\resizebox{\linewidth}{!}{
\begin{tabular}{c|c|c|c|cc}
\hline
\textbf{} & \textbf{Train} & \textbf{Eval-1} & \textbf{Eval-2} & \multicolumn{1}{c|}{\textbf{Eval-3}} & \textbf{Eval-4} \\ \hline
\# of Spks & 5,994 & 1,251 & 118 & \multicolumn{1}{c|}{816} & 816 \\ \hline
\# of Phrases & 215,432 & 23,036 & 2,310 & \multicolumn{1}{c|}{5} & 5 \\ \hline
\# of trials & \begin{tabular}[c]{@{}c@{}}TOP-1\%: 565,242\\ TOP-5\%: 903,678\\ TOP-10\%: 1,041,902\\ Random:  1,382,110\end{tabular} & \begin{tabular}[c]{@{}c@{}}TOP-1\%: 26,904\\ TOP-5\%: 52,702\\ TOP-10\%: 65,086\\ Random:  95,900\end{tabular} & \begin{tabular}[c]{@{}c@{}}TOP-10\%: 12,536\\ Random:  20,390\end{tabular} & \multicolumn{1}{c|}{21,858} & 21,733 \\ \hline
anchor & - & - & - & \multicolumn{1}{c|}{\textit{Ok Google}} & \textit{\begin{tabular}[c]{@{}c@{}}My voice is\\  my password\end{tabular}} \\ \hline
Source & Vox2-dev \cite{vox2} & Vox1 \cite{vox1} & Vox2-test \cite{vox2} & \multicolumn{2}{c}{Deepmine \cite{deepmine}} \\ \hline
\end{tabular}
}
\label{tab:vp_stats}
\end{table}

\begin{table*}[]
\centering
\renewcommand{\arraystretch}{1.35}
\caption{Main results on VoxPhrase for short-duration speaker verification, reporting EER (\%) under text-independent (TI) and text-dependent (TD) enrollment. In parentheses, d denotes the speaker embedding dimension, c denotes the frame-level feature dimension, and k indicates the number of verifier parameters.}
\resizebox{\linewidth}{!}{
\begin{tabular}{l|c|c|c|c|c|c|ccccc|ccc}
\hline
\multirow{2}{*}{\textbf{\# Exp.}} & \multirow{2}{*}{\textbf{Model}} & \multirow{2}{*}{\textbf{\# Params}} & \multirow{2}{*}{\textbf{Vox-O}} & \multirow{2}{*}{\textbf{\begin{tabular}[c]{@{}c@{}}Verifier\\ (\# Params)\end{tabular}}} & \multirow{2}{*}{\textbf{TI-Enroll}} & \multirow{2}{*}{\textbf{TD-Enroll}} & \multicolumn{5}{c|}{\textbf{Eval-1}} & \multicolumn{3}{c}{\textbf{Eval-2}} \\ \cline{8-15} 
 &  &  &  &  &  &  & \textbf{top-1\%} & \textbf{top-5\%} & \textbf{top-10\%} & \textbf{random} & \textbf{Avg.} & \textbf{top-10\%} & \textbf{random} & \textbf{Avg.} \\ \hline
E1 & \multirow{6}{*}{\begin{tabular}[c]{@{}c@{}}Ecapa-Tdnn \cite{ecapa-tdnn}\\ (d192)\end{tabular}} & \multirow{6}{*}{20.8 M} & \multirow{6}{*}{0.86} & \multirow{3}{*}{$\times$} & 10 s & $\times$ & 11.37 & 7.21 & 5.59 & 2.20 & 6.59 & 8.33 & 3.32 & 5.83 \\ \cline{1-1} \cline{6-15} 
E2 &  &  &  &  & 3 s & $\times$ & 13.51 & 9.09 & 7.28 & 3.04 & 8.23 & 10.18 & 4.29 & 7.24 \\ \cline{1-1} \cline{6-15} 
E3 &  &  &  &  & $\times$ &  \multirow{4}{*}{\begin{tabular}[c]{@{}c@{}}Phrase\\ (0.8$\sim$3 s)\end{tabular}} & 15.79 & 10.98 & 9.11 & 4.35 & 10.06 & 13.69 & 6.94 & 10.32 \\ \cline{1-1} \cline{5-6} \cline{8-15} 
E1v &  &  &  & \multirow{3}{*}{\begin{tabular}[c]{@{}c@{}}$\checkmark$\\  (c3072$\rightarrow$526 k)\end{tabular}} & 10 s &  & \textbf{9.98} & \textbf{6.30} & \textbf{4.79} & \textbf{1.94} & \textbf{5.75 (-0.84)} & \textbf{7.91} & \textbf{3.38} & \textbf{5.65 (-0.18)} \\ \cline{1-1} \cline{6-6} \cline{8-15} 
E2v &  &  &  &  & 3 s &  & 10.74 & 7.10 & 5.59 & 2.35 & 6.45 (-1.79) & 9.01 & 4.08 & 6.55 (-0.69) \\ \cline{1-1} \cline{6-6} \cline{8-15} 
E3v &  &  &  &  & $\times$ &   & 14.63 & 10.02 & 8.33 & 4.08 & 9.27 (-0.79) & 12.86 & 7.05 & 9.96 (-0.36)\\ \hline
E4 & \multirow{6}{*}{\begin{tabular}[c]{@{}c@{}}CAM++ \cite{cam++}\\ (d512)\end{tabular}} & \multirow{6}{*}{7.2 M} & \multirow{6}{*}{0.65} & \multirow{3}{*}{$\times$} & 10 s & $\times$ &  11.33 & 7.03 & 5.38 & 2.03 & 6.44 & 8.77 & 3.28 & 6.03 \\ \cline{1-1} \cline{6-15} 
E5 &  &  &  &  & 3 s & $\times$ &  13.34 & 9.03 & 7.22 & 3.01 & 8.15 & 10.75 & 4.40 & 7.58 \\ \cline{1-1} \cline{6-15} 
E6 &  &  &  &  & $\times$ &  \multirow{4}{*}{\begin{tabular}[c]{@{}c@{}}Phrase\\ (0.8$\sim$3 s)\end{tabular}} & 14.73 & 10.10 & 8.14 & 3.62 & 9.15 & 12.49 & 6.18 & 9.34 \\ \cline{1-1} \cline{5-6} \cline{8-15} 
E4v &  &  &  & \multirow{3}{*}{\begin{tabular}[c]{@{}c@{}}$\checkmark$ \\ (c512$\rightarrow$199 k)\end{tabular}} & 10 s &  & \textbf{9.58} & \textbf{5.79} & \textbf{4.44} & \textbf{1.60} & \textbf{5.35 (-1.09)} & \textbf{7.66} & \textbf{2.96} & \textbf{5.31 (-0.71)} \\ \cline{1-1} \cline{6-6} \cline{8-15} 
E5v &  &  &  &  & 3 s &  & 10.47 & 6.58 & 5.13 & 1.93 & 6.03 (-2.12) & 8.44 & 3.48 & 5.96 (-1.62) \\ \cline{1-1} \cline{6-6} \cline{8-15} 
E6v &  &  &  &  & $\times$ &   & 13.59 & 9.17 & 7.39 & 3.09 & 8.31 (-0.84) & 11.77 & 5.42 & 8.60  (0.74)\\ \hline
E7 & \multirow{6}{*}{\begin{tabular}[c]{@{}c@{}}ERes2Net-L \cite{eres2net} \\ (d192)\end{tabular}} & \multirow{6}{*}{20.5 M} & \multirow{6}{*}{0.57} & \multirow{3}{*}{$\times$} & 10 s & $\times$ &  9.32 & 5.73 & 4.44 & 1.59 & 5.27 & 6.70 & 2.57 & 4.64 \\ \cline{1-1} \cline{6-15} 
E8 &  &  &  &  & 3 s & $\times$ &  11.02 & 7.17 & 5.64 & 2.21 & 6.51 & 8.18 & 3.29 & 5.74 \\ \cline{1-1} \cline{6-15} 
E9 &  &  &  &  & $\times$ &  \multirow{4}{*}{\begin{tabular}[c]{@{}c@{}}Phrase\\ (0.8$\sim$3 s)\end{tabular}} & 12.92 & 8.81 & 7.10 & 2.99 & 7.96 & 11.15 & 4.84 & 8.00 \\ \cline{1-1} \cline{5-6} \cline{8-15} 
E7v &  &  &  & \multirow{3}{*}{\begin{tabular}[c]{@{}c@{}}$\checkmark$ \\ (c1024$\rightarrow$264 k)\end{tabular}} & 10 s &  & \textbf{8.17} & \textbf{4.89} & \textbf{3.74} & \textbf{1.37} & \textbf{4.54 (-0.73)} & \textbf{6.50} & \textbf{2.50} & \textbf{4.50 (-0.14)} \\ \cline{1-1} \cline{6-6} \cline{8-15} 
E8v &  &  &  &  & 3 s &  & 8.99 & 5.57 & 4.30 & 1.64 & 5.13 (-1.39) & 7.26 & 3.09 & 5.18 (-0.56) \\ \cline{1-1} \cline{6-6} \cline{8-15} 
E9v &  &  &  &  & $\times$ &   & 11.94 & 7.91 & 6.36 & 2.68 & 7.22 (-0.73) & 10.26 & 4.79 & 7.53  (-0.47)\\ \hline
\end{tabular}
}
\label{tab:main_result}
\end{table*}

\section{Results}

\subsection{Comparative Evaluation on VoxPhrase}

Table~\ref{tab:main_result} presents the main results on VoxPhrase using different enrollment strategies. Under the evaluated settings (3 s and 10 s TI enrollment versus 0.8–3 s TD enrollment), TI enrollment consistently outperforms TD across all models. This indicates that, within the tested duration range, the stability advantage brought by longer enrollment duration outweighs the benefits of phonetic consistency provided by TD enrollment. While TD ensures lexical alignment between enrollment and query speech, its limited duration constrains the stability of speaker representations. In contrast, TI enrollment benefits from richer speaker-discriminative information, leading to lower error rates. Additionally, hard trials are particularly prominent in short-duration settings, resulting in increased false rejection rates.

Introducing the neural verifier in the TD scenario results in consistent performance improvements across all models, highlighting the importance of frame-level interaction modeling for short-duration speaker verification. Hybrid enrollment combined with neural re-scoring yields significant performance improvements over both pure TI and pure TD enrollment, demonstrating that the proposed method effectively integrates both types of enrollment information, significantly enhancing short-duration speaker verification performance. Notably, in hard-case scenarios, the hybrid method continues to deliver substantial gains, showcasing its robustness in challenging conditions.

\subsection{Effect of TI Duration}

The TD enrollment achieves an EER of 3.62\%, which improves to 3.09\% with the verifier, confirming the effectiveness of frame-level modeling. In contrast, TI enrollment with a duration of 3s performs poorly (8.86\%). As the TI duration increases from 1s to 10s, the EER continuously decreases. When $T > 3s$, TI outperforms TD, but when $T < 2s$, TI remains inferior to TD, highlighting the impact of TI enrollment on speaker representation under short-duration conditions.

\begin{figure}[]
  \centering
  \includegraphics[width=\linewidth]{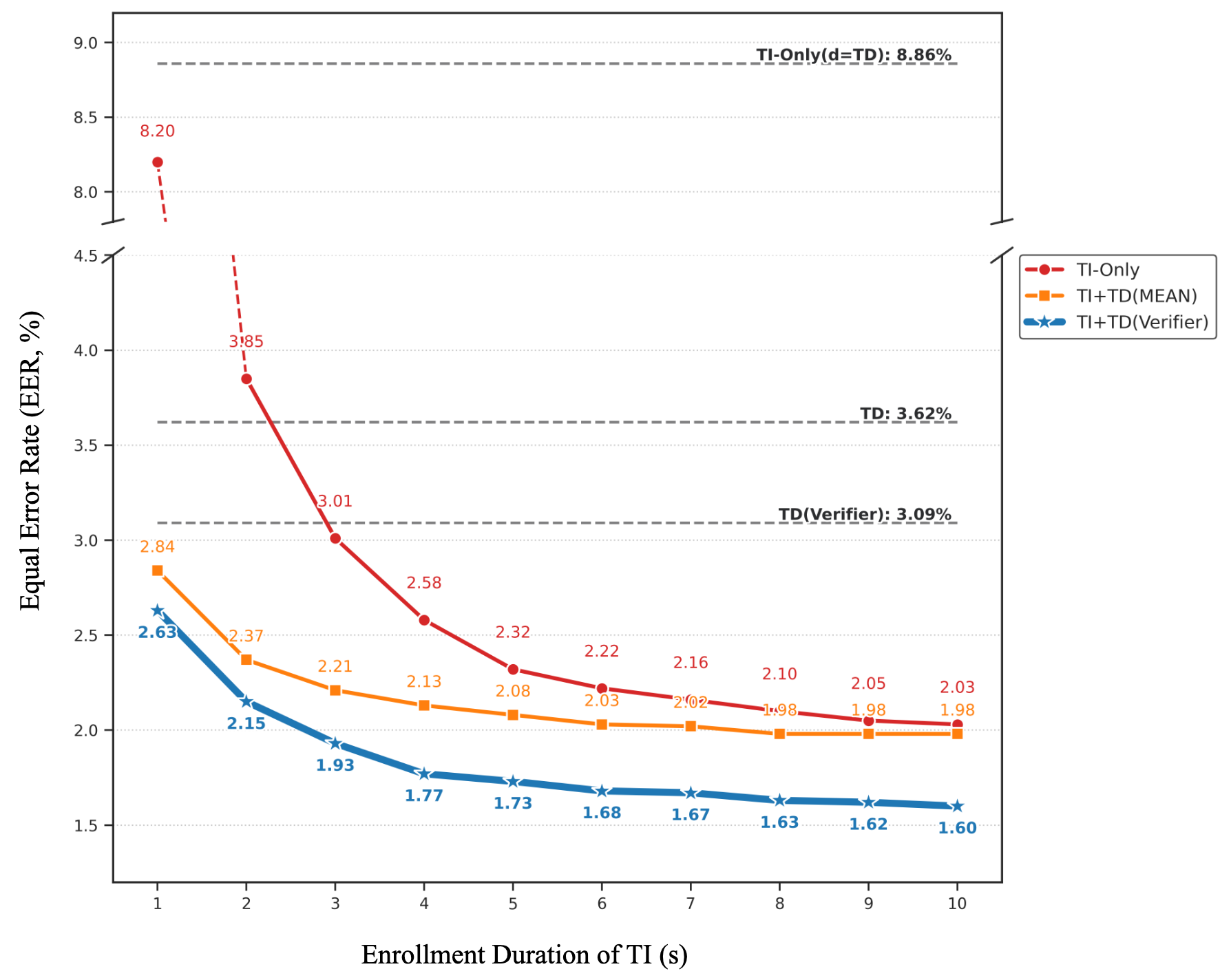}
  \caption{The effect of TI enrollment duration on EER under different enrollment conditions, evaluated on Eval-1 (random) setting.}
  \label{fig:svm2}
\end{figure}

The TI+TD (mean) baseline also improves with longer TI durations but gradually converges to TI-only performance (2.03\% vs. 1.98\% at 10s). In comparison, the proposed hybrid enrollment with neural re-scoring consistently achieves the best results, reaching 1.6\% at 10s, demonstrating its effectiveness and robustness.

\begin{table}[]
\centering
\renewcommand{\arraystretch}{1.35}
\caption{Performance on Out-of-Distribution evaluation.}
\resizebox{\linewidth}{!}{
\begin{tabular}{c|c|c|c|c|c}
\hline
\textbf{Model} & \textbf{Verifier} & \textbf{TI-Enroll} & \textbf{TD-Enroll} & \textbf{Eval-3} & \textbf{Eval-4} \\ \hline
\multirow{4}{*}{\begin{tabular}[c]{@{}c@{}}CAM++ \cite{cam++} \\ (d512)\end{tabular}} & \multirow{2}{*}{$\times$} & $\times$ & $\checkmark$ & 8.17 & 6.19 \\ \cline{3-6} 
 &  & $\checkmark$ & $\times$ & 10.96 & 4.91 \\ \cline{2-6} 
 & \multirow{2}{*}{$\checkmark$} & $\times$ & $\checkmark$ & 7.72 & 4.62 \\ \cline{3-6} 
 &  & $\checkmark$ & $\checkmark$ & \textbf{6.71} & \textbf{3.48} \\ \hline
\multirow{4}{*}{\begin{tabular}[c]{@{}c@{}}ERes2Net-L \cite{eres2net} \\ (d192)\end{tabular}} & \multirow{2}{*}{$\times$} & $\times$ & $\checkmark$ & 6.97 & 4.54 \\ \cline{3-6} 
 &  & $\checkmark$ & $\times$ & 8.86 & 3.58 \\ \cline{2-6} 
 & \multirow{2}{*}{$\checkmark$} & $\times$ & $\checkmark$ & 6.72 & 3.22 \\ \cline{3-6} 
 &  & $\checkmark$ & $\checkmark$ & \textbf{4.88} & \textbf{2.38} \\ \hline
\end{tabular}
}
\end{table}

\subsection{Performance on Out-of-Distribution Data}

We analyzed the performance of the trained verifier on out-of-distribution data using the Deepmine \cite{deepmine} dataset. The results show that TD enrollment outperforms TI enrollment, where TI enrollment is based on short-duration audio. TD enrollment provides higher stability and effectiveness through lexical alignment. Additionally, the proposed hybrid enrollment method, combining TD and TI features with neural re-scoring, significantly improves performance, confirming its effectiveness and robustness in short-duration speaker verification.

\section{Conclusion}

In this work, we proposed a hybrid enrollment method combining text-dependent (TD) and text-independent (TI) features with neural re-scoring to improve short-duration speaker verification. Our experiments on the VoxPhrase dataset show that, under practical enrollment durations (3 s and above), TI enrollment consistently outperforms TD enrollment by providing more stable and comprehensive speaker representations. However, further analysis reveals a duration-dependent trend: when enrollment duration becomes extremely short, the lexical alignment advantage of TD enrollment becomes more pronounced. The proposed hybrid method effectively integrates the complementary strengths of both TD and TI, achieving consistent improvements across backbone models and evaluation conditions, particularly in hard-case and out-of-distribution scenarios. This demonstrates the robustness and effectiveness of the hybrid approach for real-world short-duration speaker verification tasks.

\newpage

\section{Acknowledgments}
This work was supported in part by the 6G Science and Technology Innovation and Future Industry Cultivation Special Project of Shanghai Municipal Science and Technology Commission under Grant 24DP1501001, in part by the National High Quality Program under Grant TC220H07D, and in part by the Xi'an Jiaotong-Liverpool University under Grant for ILAI.

\section{Generative AI Use Disclosure}
Generative AI tools were used in this work solely for language editing, polishing, and formatting of the manuscript. They were not used to generate any core content, research ideas, experimental designs, results, or substantial portions of the text. All scientific contributions—including model design, experimentation, analysis, and conclusions—were carried out entirely by the authors.

\bibliographystyle{IEEEtran}
\bibliography{mybib}

\end{document}